# One-dimensional polarization-hybrid photonic crystal molecules

*Tiantong Li and Katia Gallo*[*]

Nonlinear and Quantum Photonics, Department of Physics, KTH - Royal Institute of Technology, Roslagstullsbacken 21, 114 21 Stockholm, Sweden.

* Corresponding author Katia Gallo. e-mail : gallo@kth.se

**ABSTRACT**

Photonic molecules, i.e. artificial structures composed of coherently coupled optical cavities, are paradigmatic systems for investigating fundamental phenomena across photonics, quantum optics and topological physics. In recent years, photonic integrated circuits have emerged as a particularly powerful platform for their realization, exploiting also additional synthetic dimensions afforded by the degrees of freedom of light. To date, however, photonic molecule implementations have relied almost entirely on geometries defined by spatial coupling and lattice symmetries rather than polarization.

Here, we introduce a fundamentally new class of photonic molecules in which polarization is exploited as the primary dimension in the device response. By harnessing fundamental guided-mode couplings sustained by engineered Bragg gratings in photonic waveguides, we establish a new paradigm to access in 1D formats the coupled-resonator physics traditionally associated with higher-dimensional or free-space systems, demonstrating prototypical devices which can support Fano resonances or resonance-splitting for signals in the telecom band. Besides corroborating the theoretical predictions, experimental realizations in thin film lithium niobate open new prospects for the further exploration of novel reconfigurable topological, non-Hermitian and quantum photonic circuits, relying on the intrinsic nonlinear and electro-optic functionalities of this platform.

## Introduction

Coupled optical cavities provide direct access to coherent phenomena underpinning a wide range of physical effects, from bound state in the continuum, topological band formation and exceptional points, to a broad spectrum of quantum and nonlinear optical phenomena,[1-3] which often ultimately find their roots in coupled dual-resonator models corresponding to photonic molecules (PM).[4-9] Despite their relative simplicity, such prototypical systems exhibit a very rich interaction landscape, providing all-optical analogues of atomic systems which can span the full range from Fano interference (FI) to Auler-Townes splitting (ATS) in strong coupling regimes.[8-11] Furthermore, their realization in photonic integrated circuit (PIC) architectures, made possible by recent developments in platforms such as silicon, silicon nitride and lithium niobate on insulator (LNOI), affords refined control of mode confinement, phase, and coupling strengths, while providing remarkable footprint reduction, robustness and scalability, key for fundamental as well as applied science breakthroughs.[5,12-18]

A straightforward way to implement coupled resonances is to fabricate two physical cavities that are coupled through evanescent fields. However, accessing photonics degrees of freedom that go beyond the spatial dimensions may unlock further functionalities and enable brand new device concepts in synthetic spaces,[19-21] as recently demonstrated with multimode resonators and frequency crystals.[22-24] However, the synthetic space afforded by photon polarization remains still essentially unexploited for PM functionalities, due to challenges in its manipulation and control in PIC geometries, which typically imply the use of strongly birefringent materials, periodic poling, or careful waveguide dispersion engineering combined with high-order mode conversion, requiring complex technological processes and/or PIC architectures.[25-29]

Here we design and demonstrate a novel class of integrated PMs that leverage the polarization degree of freedom in monolithic 1D cavity formats, using integrated Bragg grating technology. On this compact and versatile platform, we achieve excellent agreement between theory and experiments realizing two fundamental operation principles for PMs, namely mode hybridization and pathway interference, engineerable by grating design, which clearly manifest themselves in the device spectral response through ATS and FI features. The approach underpins a new class of polarization-enabled photonic crystal molecules compatible with most high-confinement PIC platforms. Moreover, the experimental realization in a PIC platform such as LNOI,[30,31] paves the way for a brand new class of low-footprint devices affording novel programmable nonlinear and quantum functionalities,[5,32,33] as well as fundamental explorations of non-Hermitian and topological physics in high-dimensional spaces, leveraging polarization control and engineering in combination with the electro-optic and all-optical functionalities of this unique PIC platform.

## Results

We start by describing the principle of operation, design and testing of the novel integrated component enabling the target polarization-control functionalities, consisting of a distributed Bragg reflector (DBR) tailored to couple fundamental modes of orthogonal polarization (quasi $TE_{00}$ and $TM_{00}$) in a counter-propagating fashion. We then present the results obtained by combining two such DBRs (for simplicity identical) to create Fabry-Perot cavities, which support polarization-maintaining ($TE_{00}$-$TE_{00}$ or $TM_{00}$-$TM_{00}$) as well as hybridized ($TE_{00}$-$TM_{00}$) resonant modes, realizing photonic molecules in the synthetic dimension offered by the polarization degree of freedom. Finally, we show with theory and experiments on the LNOI platform how the interplay of the above processes can be engineered by suitable design of the DBRs to access either the ATS regime or FI in the telecom band in integrated 1D cavity devices.

### Polarization-coupling reflectors

The key building block for the functional devices considered here consists of a polarization-coupling DBR designed to operate in the telecom band with the fundamental ($TE_{00}$ and $TM_{00}$) modes of a sidewall-modulated photonic rib waveguide, as sketched in *Figure 1a*. The integrated DBR section features a single-sided modulation of the rib top width between the values $w_1$ and $w_2$ (along the transverse axis $x$), and a constant period $\Lambda$ and 50% duty cycle (along the longitudinal axis $z$). In contrast to previous architectures,[34] the chosen asymmetric waveguide modulation realizes a contra-directional coupling between fundamental guided modes of orthogonal polarization while allowing at the same time the polarization-preserving Bragg coupling features of a conventional DBR.[35,36] This is illustrated by *Fig. 1a* for an incident forward-propagating $TE_{00}^+$ field producing the backward-reflected $TE_{00}^-$ and $TM_{00}^-$ waves. All the possible counter-propagating couplings enabled by the DBR grating are summarized by the diagram in *Fig. 1b*, where $\gamma$, $\kappa_{TE}$, and $\kappa_{TM}$ designate the coupling coefficients of the polarization-converting ($TE_{00}$-$TM_{00}$), polarization-maintaining TE ($TE_{00}$-$TE_{00}$) and TM ($TM_{00}$-$TM_{00}$) interactions, respectively.

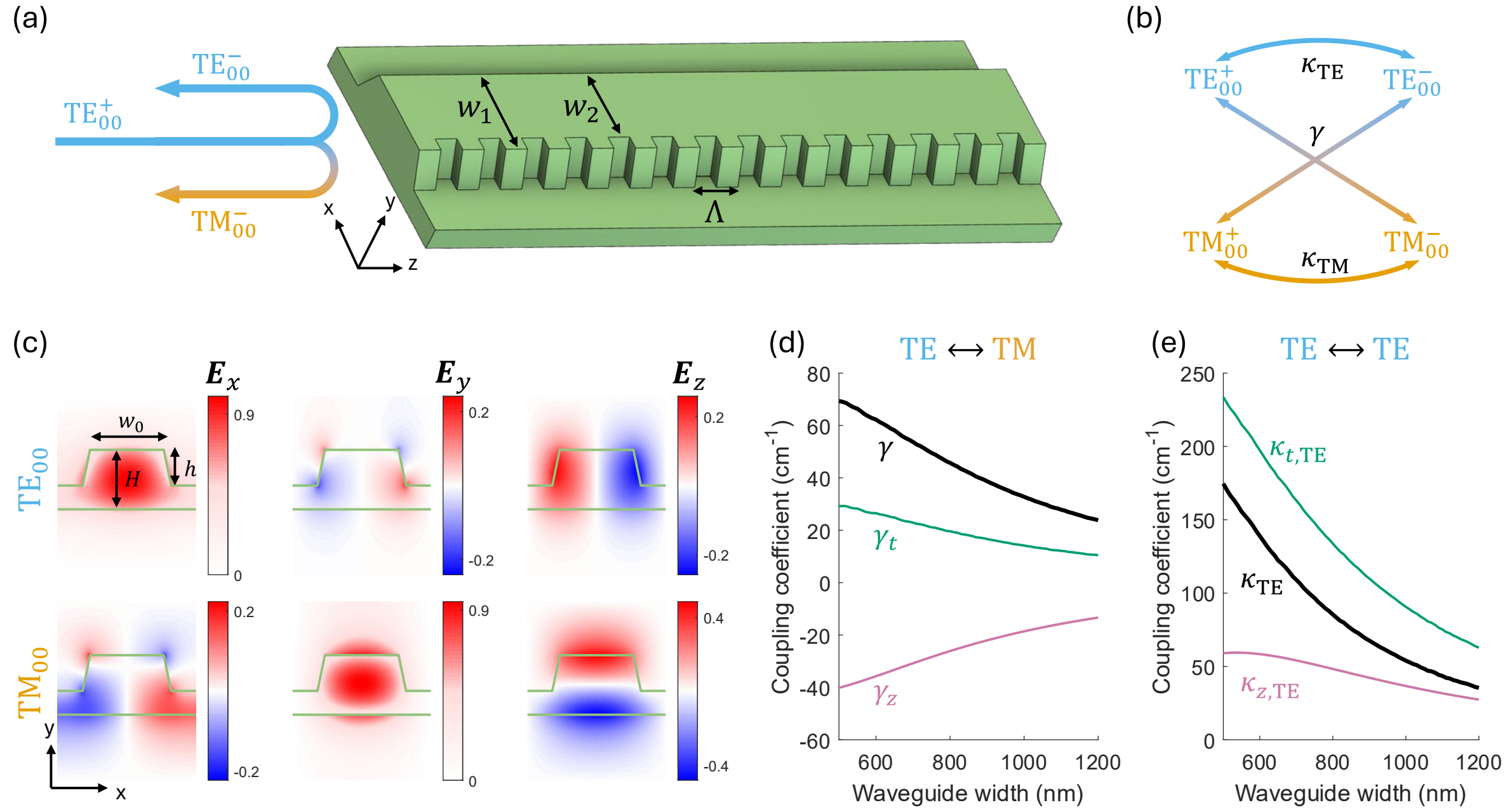

**Figure 1**. Principle of operation of a polarization-coupling DBR. **(a)** Schematic of the polarization conversion Bragg grating on X-cut TFLN. The light propagates along the crystalline Y-axis of lithium niobate. The + (–) sign denotes forward (backward) waves. **(b)** Coupling diagram for the fundamental TE and TM modes with corresponding coupling coefficients. **(c)** Transverse distributions of the vectorial electric field components ($\boldsymbol{E_x}$, $\boldsymbol{E_y}$, $\boldsymbol{E_z}$) for the $TE_{00}$ and $TM_{00}$ mode computed for a TFLN rib waveguide (green lines) with: $w_0$ = 800 nm, $h$ = 300 nm, $H$ = 500 nm. Computed coupling coefficients for: **(d)** $TE_{00}$-$TM_{00}$ and **(e)** and $TE_{00}$- $TE_{00}$ modal interactions as a function of waveguide width $w_0$, for a sidewall modulation depth $\delta w = w_1 - w_2$ = 100 nm. Green and purple lines: transverse and longitudinal components of the total coupling strength (black lines).

The device response is analyzed by coupled mode theory (CMT) computations based on *Equation S22-25 in Supplementary Information*, with key parameters obtained from eigenmode analyses of the unperturbed waveguide structures, performed with a commercial vectorial solver (MODE, Ansys Lumerical). As an example, referring to the technology platform used for the experiments, *Fig. 1c* shows the computed transverse field profiles of the guided modes at a wavelength $\lambda$ = 1550 nm for a rib waveguide of top width $w_0$ = 800 nm, height $h$ = 300 nm and sidewall angle $\theta$ = 75° in an X-cut 500nm-thin film of congruent $LiNbO_3$.[37] The simulations make apparent how the tight optical confinement in the rib waveguide results in strong longitudinal fields ($\boldsymbol{E_z}$) in addition to the conventional transverse ($\boldsymbol{E_t}$) components of the $TE_{00}$ and $TM_{00}$ modes, which are preferentially polarized along $x$ and $y$, respectively, in the chosen frame of reference (*Fig. 1c*). As indicated by the scale bars in each plot, the longitudinal and transverse fields exhibit comparable amplitudes, with the ratio between the maxima of $\boldsymbol{E_z}$ and $\boldsymbol{E_t}$ amounting to 26% for the $TE_{00}$ mode and 52% for the $TM_{00}$ mode. The presence of significant longitudinal field components in combination with the chosen asymmetric perturbation of the waveguide is at the origin of a non-zero polarization-mixing, occurring through an overall TE-TM coupling coefficient $\gamma = \gamma_t - \gamma_z$, which stems from the combination of a transverse and a longitudinal contribution, given by:

$$\gamma_t = \frac{\omega}{4}\iint \epsilon_1(x,y)\, \boldsymbol{E}_t^{(\mathrm{TE})}(x,y) \cdot \boldsymbol{E}_t^{(\mathrm{TM})^*}(x,y)\, \mathrm{d}x\mathrm{d}y$$
$$\gamma_z = \frac{\omega}{4}\iint \epsilon_1'(x,y)\, \boldsymbol{E}_z^{(\mathrm{TE})}(x,y) \cdot \boldsymbol{E}_z^{(\mathrm{TM})^*}(x,y)\, \mathrm{d}x\mathrm{d}y \qquad (1)$$

respectively. Here $\omega$ is the frequency of light, while $\epsilon_1$ and $\epsilon_1'$ are the first order Fourier coefficients of the periodic index perturbation seen by the transverse and longitudinal fields, respectively (see also *Supplementary Information S1*). The computed values of the overall TE-TM coupling strength $\gamma$ (black line) are plotted in *Fig. 1d* together with the transverse and longitudinal couplings $\gamma_t$ and $\gamma_z$ (green and purple lines), respectively, for different waveguide widths $w_0$. The figure clearly shows a significant enhancement in the coupling stemming from the longitudinal term, when compared to the transverse term alone. The coupling coefficients for conventional, polarization-maintaining TE and TM interactions in the DBR structure are determined in a similar fashion, considering the modal distributions of the appropriate field components (details in *Supplementary Information S1*). As an example, *Fig. 1e* plots the $\mathrm{TE}_{00}$-$\mathrm{TE}_{00}$ Bragg coupling coefficient $\kappa_{\mathrm{TE}}$ as a function of $w_0$, together with its individual components $\kappa_{t,\mathrm{TE}}$ and $\kappa_{z,\mathrm{TE}}$, computed for the same structure as in *Fig. 1d*, revealing here a dominant role of the transverse coupling.

Phase-matching plays an equally critical role in controlling the inter-modal power transfer and determining the spectral response of the DBR. The intended spectral working points in the 1550-1600 nm telecom band for the LNOI integrated devices are selected by suitable choice of the DBR grating period $\Lambda$, as illustrated by *Fig. 2a*, where we show the dispersion of the $\mathrm{TE}_{00}$ and $\mathrm{TM}_{00}$ guided mode propagation constants, $\beta_{\mathrm{TE}}(\lambda)$ and $\beta_{\mathrm{TM}}(\lambda)$ computed for the waveguide geometry of *Fig 1c* ($w_0$ = 800 nm, $h$ = 300 nm, $H$ = 500 nm). The horizontal dashed line drawn in the same diagram at $\pi/\Lambda$ highlights the Bragg resonance condition with a period $\Lambda$ = 435 nm. Its crossing points with the $\mathrm{TM}_{00}$ and $\mathrm{TE}_{00}$ dispersion curves, at $\lambda_{\mathrm{TM}}$ = 1573.6 nm and $\lambda_{\mathrm{TE}}$ = 1593.8 nm, respectively, mark the phase-matching wavelengths for polarization-maintaining DBR operation. On the other hand, the polarization-conversion wavelength $\lambda_C$ = 1583.4 nm, highlighted by the circle on the same line in *Fig 2a*, marks the phase-matching condition whereby a $\mathrm{TE}_{00}$ input is converted to a $\mathrm{TM}_{00}$ reflected output (and vice versa), occurring when the mismatch $\Delta\beta_C = \beta_{\mathrm{TE}} + \beta_{\mathrm{TM}} - 2\pi/\Lambda = 0$.

Further insights into the expected spectral response of the DBR are gained by numerically solving the coupled mode equations detailed in *Supplementary Information S1.* The orange (blue) line in *Fig. 2b* shows the computed TM (TE) transmission for a DBR of length $L_B$ = 900 and period $\Lambda$ = 391.5 µm, As intuitively expected, both polarizations feature identical photonic bandgaps centered at the polarization-coupling wavelength $\lambda_C$ in addition to their – purely TM or TE – bandgaps, located at their respective Bragg resonances ($\lambda_{TM}$ or $\lambda_{TE}$), in agreement with phase matching considerations.

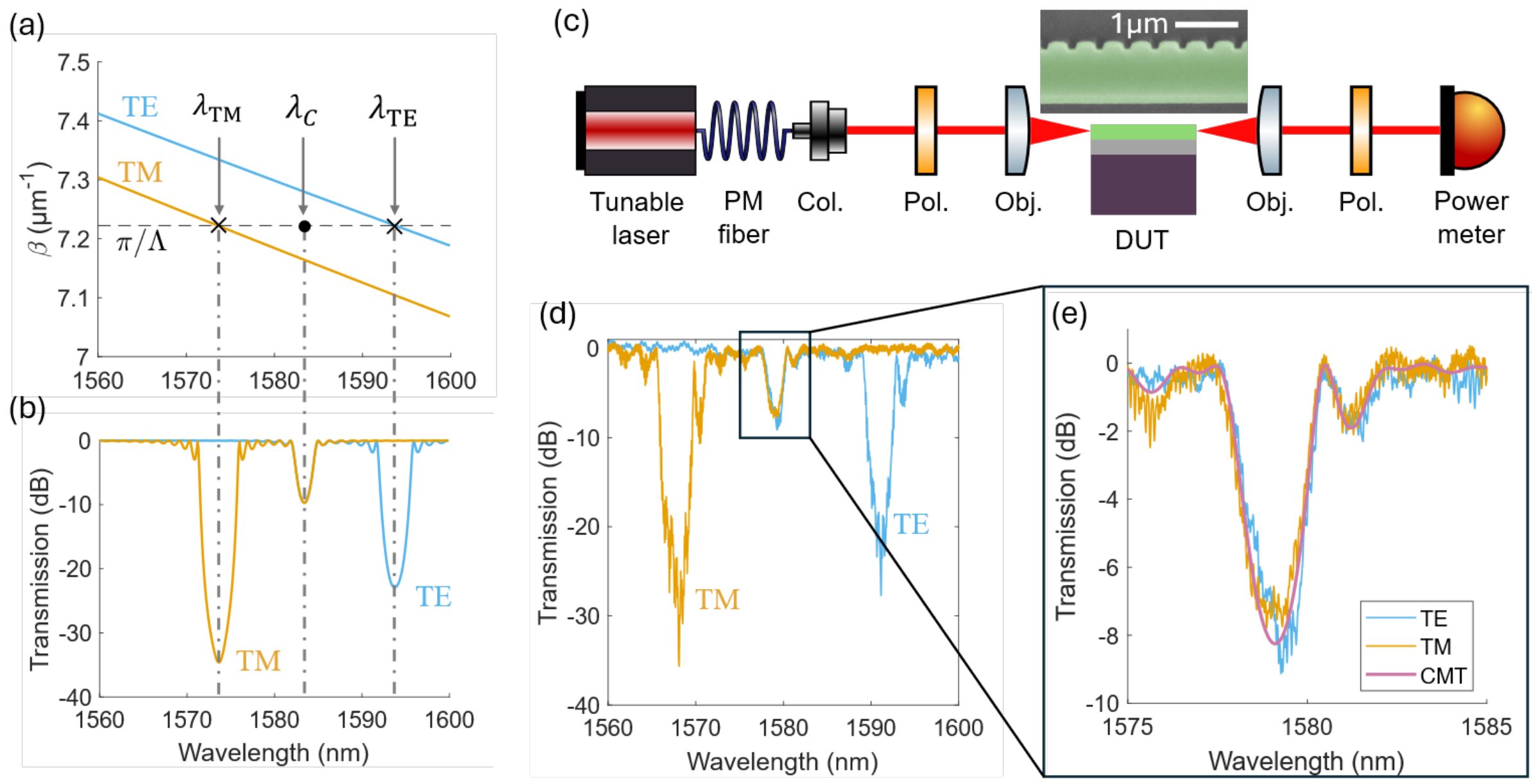

**Figure 2.** Spectral response of a polarization coupling DBR operating the telecom band. **(a)** Plot of the TE$_{00}$ ($\beta_{TE}$) and TM$_{00}$ ($\beta_{TM}$) mode propagation constants as a function of wavelength for the DBR of Fig.1 and a grating period $\Lambda$ = 435 nm. The horizontal dashed line corresponds to $\pi/\Lambda$. $\lambda_C$ = 1583.4 nm, $\lambda_{TM}$ and $\lambda_{TE}$ mark the phase matching wavelengths for polarization conversion, and polarization-maintaining TM (orange) and TE (blue) Bragg reflection, respectively. **(b)** Computed TE (orange line) and TM (blue line) transmission spectra of a DBR of length $L_B$ = 900$\Lambda$. **(c)** Experimental setup for polarization-resolved transmission measurements. Col. = collimator; Pol.= linear polarizer; Obj.= objective; DUT = device under test. Insert: scanning electron microscopy image of a fabricated DBR. **(d)** Measured TE and TM transmission spectra of a fabricated 900$\Lambda$-long DBR. **(e)** Zoomed-in view of the polarization conversion band, comparing TE and TM measurements (orange and blue curves) and CMT simulations (purple line).

Design guidelines for DBR demonstrators on the X-cut 500nm-thick LNOI platform were developed from CMT simulations as those of *Fig. 2b*. Details of the fabrication process and optical measurements can be found in the *Methods* section. The setup of *Fig. 2c*, employing a narrow-linewidth tunable telecom source in combination with a pair of free-space polarizers (placed at the waveguide input and output, respectively) was used to experimentally control and track the TE and TM polarization for a direct comparison with the theoretical predictions. As an example, *Fig.2b* shows the transmission spectra obtained for a DBR device with $\Lambda$ = 435 nm, $L_B$ = 900$\Lambda$ and a sidewall modulation amplitude $\delta w = w_1 - w_2$ = 100 nm. The experiments match well the theoretical predictions, yielding three distinct reflection bands, located at $\lambda_{TM}$ =

1568 nm, $\lambda_C$ = 1579 nm and $\lambda_{TE}$ = 1591 nm, corresponding to the expected adjacent resonances of the device (*Fig. 2a*). When measured in either the TE or TM polarization, the hybrid (TE-TM) photonic bandgap exhibits an identical profile, as seen in the zoomed-in plot of *Fig. 2e*, proving that polarization conversion is happening as expected inside this band. The very good agreement with theory is further highlighted by the solid line overlayed to the experimental data in the same plot, showing the result of simulations of the device transmission, accounting also for fabrication imperfections. With a polarization conversion efficiency of 86.5% over a 391.5 μm DBR, the TE-TM coupling coefficient is estimated to be $\gamma$ = 43 cm$^{-1}$, similar to the prediction from *Eq. 1*. Zoomed-in views of the two polarization-maintaining bandgaps are shown in *Supplementary Information S3* and feature equally good agreement between theory and experiments, with coupling strengths of $\kappa_{TE}$ = 82 cm$^{-1}$ and, $\kappa_{TM}$ = 110 cm$^{-1}$.

After validating the polarization-coupling functionalities of the DBR device and verifying the predictive capabilities of our CMT model, Fabry-Perot cavity designs were developed for elementary photonic molecules defined by pairs of identical single-sidewall DBR gratings of period $\Lambda$ and length $L_B$, separated by a uniform section of length $\delta L \geq \Lambda$ (as sketched in *Fig. 3b* and *4b*). This provided the prototypical system for ultralow-footprint 1D photonic molecules operating in the polarization synthetic space, which we then implemented on the LNOI integrated platform.

## Photonic molecule: strong coupling regime

The strong coupling ATS regime is achieved with suitably optimized DBR gratings at the polarization-hybrid TE-TM resonance ($\lambda_C$ in *Fig. 2*), where the reflectors operate as polarization converters. The 1D cavity then supports two polarization-hybrid resonances: one involving a forward-propagating TE and backward TM mode, denoted as [$TE^+$ - $TM^-$] and the other, involving the complementary forward-TM and backward-TE mode [$TM^+$ - $TE^-$]. The two resonances are only differentiated by the propagation direction and inherently spectrally degenerate. Any mechanism providing strong coupling between the two cavity modes can break the degeneracy and result in resonance-splitting.[38] The coupling in our case is provided by the spectrally-adjacent polarization-preserving resonances of DBRs, occurring at the wavelengths $\lambda_{TE}$ and $\lambda_{TM}$. Accordingly, the ensuing resonance splitting of the polarization-hybrid cavity mode can be engineered by fine-tuning the length ($L_B$) and TE and TM Bragg coupling strengths ($\kappa_{TE}$ and $\kappa_{TM}$) of the DBRs. The full set of interactions underpinning the ATS cavity response at $\lambda_C$ are summarized in the diagram of *Fig. 3a*, where solid (dashed) lines indicate phase-matched (partially mismatched) couplings.

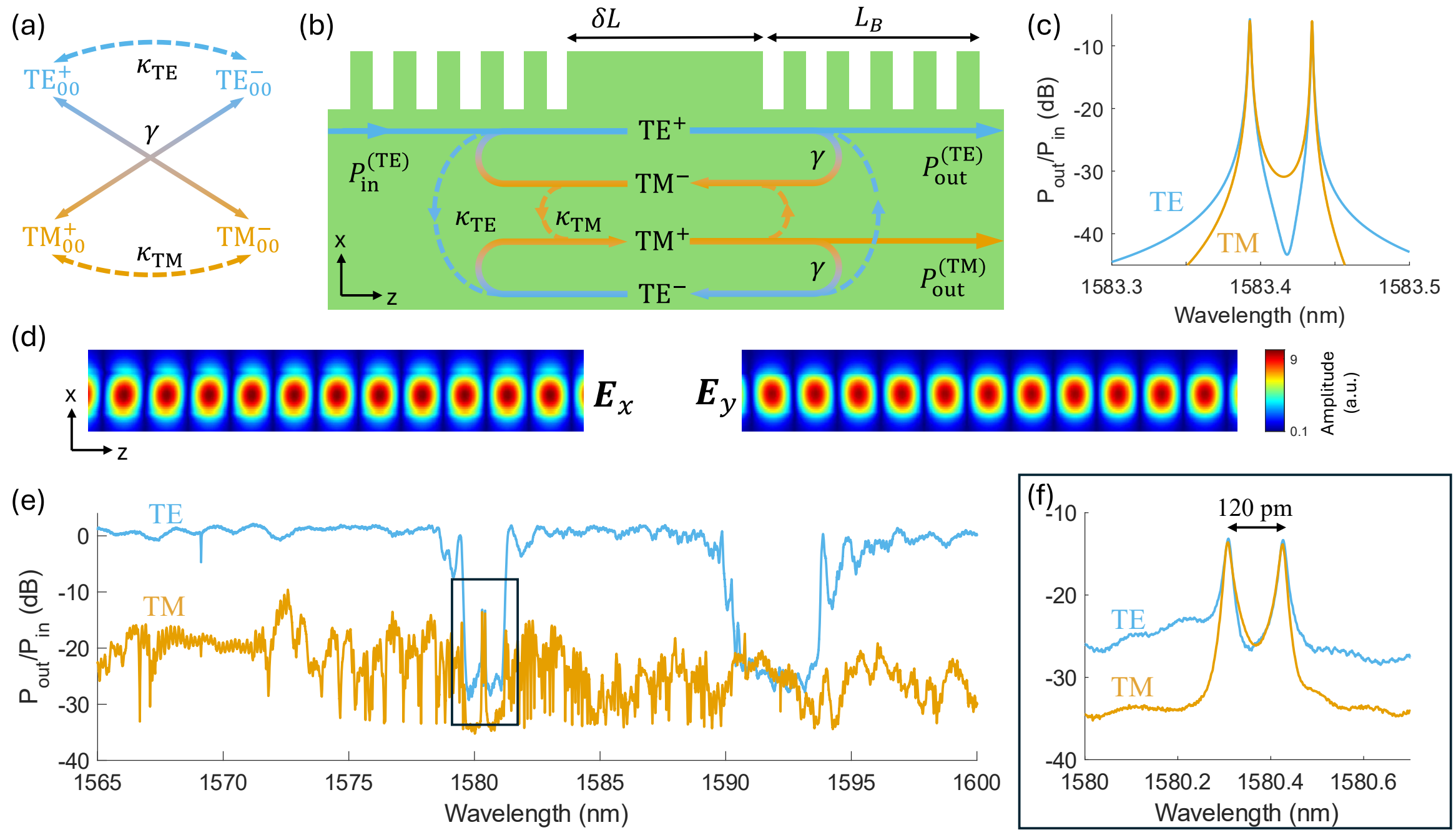

**Figure 3**. Polarization-hybrid photonic-molecule in the the strong coupling ATS regime. **(a)** Modal coupling diagram at $\lambda_C$. Solid arrows: phase-matched TE-TM coupling (γ). Dashed arrows: phase-mismatched TE-TE ($\kappa_{TE}$) and TM-TM ($\kappa_{TM}$) coupling. **(b)** Schematic of the relevant interactions for a cavity and DBRs of lengths $\delta L$ and $L_B$, respectively, excited with a power $P_{in}^{(TE)}$ in the forward-propagating $TE_{00}$ mode. **(c)** TE (blue) and TM (orange) output powers, normalized to $P_{in}^{(TE)}$), obtained by CMT analysis for the DBR of Fig. 2 with $\delta L = \Lambda$, $L_B$= 200 $\Lambda$. **(d)** Propagation plots (*z*-*x*) of the TE and TM vectorial field components $\boldsymbol{E}_x$ and $\boldsymbol{E}_y$ at the center of a 100 µm-long cavity, at resonance. **(e)** Spectral distributions of the output TE (blue) and TM (orange) components measured for a cavity with $L_B = 1750\Lambda$ and $\delta L = \Lambda$. **(f)** Zoomed-in view of the TE-TM resonances around $\lambda_C$.

*Fig. 3b* illustrates the device principle of operation, considering an incident forward-propagating $TE^+$ mode, which is first reflected by the second mirror as a backward $TM^-$ mode, thus seeding the polarization-hybrid [$TE^+$-$TM^-$] cavity resonance. The DBR gratings, however, provide also weaker polarization-maintaining couplings between counterpropagating TM and TE modes (via $\kappa_{TM}$ and $\kappa_{TE}$), thus feeding power into the complementary [$TM^+$-$TE^-$] cavity mode and coupling the latter to the [$TE^+$-$TM^-$] mode directly excited by the $TE^+$ pump. The cavity interactions at $\lambda_C$ are analyzed quantitatively by a CMT and transfer-matrix method developed for phase-shifted Bragg grating cavities (see also *Supplementary Information S1-2*),[35] yielding the results illustrated by *Fig. 3c*. There we plot the computed TE and TM output powers [$P_{out}^{(TE)}$ in blue and $P_{out}^{(TM)}$ in orange, respectively], normalized to the purely-TE input [$P_{in}^{(TE)}$], as a function of wavelength around the polarization-hybrid resonance at $\lambda_C$ = 1583.4 nm (cf *Fig. 2b*). The simulations clearly exhibit a spectral splitting of the cavity resonance, apparent in both the TE and TM outputs, with two peaks corresponding to the symmetric and anti-symmetric mixture of the [$TE^+$-$TM^-$] and [$TM^+$-$TE^-$] cavity modes. The linewidth of each peak ($\delta\lambda$) is controlled by the coupling strength $\gamma$ and the grating length $L_B$, while their spectral separation ($\Delta\lambda$) is controlled by the coupling strengths $\kappa_{TE}$ and $\kappa_{TM}$ and the associated phase mismatches $\Delta\beta_{TE}$ and $\Delta\beta_{TM}$ at $\lambda_C$. The resonance splitting is resolved when the linewidth $\delta\lambda$ is small enough to access a strong coupling regime,[11] which is clearly the case in *Fig 3c* for a DBR length $L_B$ = 870 μm, yielding $\delta\lambda$ = 1.5 pm and $\Delta\lambda$ = 42 pm.

The CMT predictions were additionally validated by independent simulations performed with an eigenmode expansion method (Ansys Lumerical), without relying on the perturbative CMT approach based on fundamental guided-mode coupling. The numerical results, detailed in *Supplementary Information S4*, exhibit very good agreement with the CMT predictions, confirming that the latter is effectively capturing all the relevant physics underlying the device spectral response (*Fig. 3c*). Moreover, the simulations using the eigenmode expansion method allow to retrieve the full vectorial electric field distribution inside the cavity. This is shown in *Fig. 3d* at the wavelength corresponding to one of the peaks in the splitting, with color maps for the amplitude of the $\boldsymbol{E_x}$ and $\boldsymbol{E_y}$ transverse field components, for a 10 μm-long central segment in a nanowire cavity with DBR separation $\delta L$ = 100 μm (top view, *x*-*z* plane). The standing wave patterns seen in both the $\boldsymbol{E_x}$ (TE) and $\boldsymbol{E_y}$ (TM) distributions further confirm the hybrid cavity mode coupling occurring around $\lambda_C$.

Experimental evidence for the theoretically predicted splitting (*Fig. 3c*) was provided by measurements on integrated Fabry-Perot cavities implemented on the LNOI platform. In *Fig. 3e* we show the TE and TM output spectra obtained from an integrated cavity with $w_1$ = 750 nm, $w_2$ = 850 nm, $\Lambda = \delta L$ = 435 nm, and $L_B = 1750\Lambda$ = 761.25μm, with pure TE input excitation. For a direct comparison with the simulations of *Fig. 3c*, in *Fig. 3f* we show also the details of the spectral response experimentally recorded at the center of the TE-TM bandgap in the same device, making apparent a resonance splitting of $\Delta\lambda$ = 120 pm. The full-width-half-maximum of the two, nearly identical, resonance peaks featured by the TE and the TM responses are = 16 pm, corresponding to a quality factor of $0.988\times10^5$.

The experimental data of *Fig. 3e* bear also the signatures of the TE and TM photonic bandgaps centered around $\lambda_{TE}$ = 1592 nm and $\lambda_{TM}$ = 1569 nm, respectively. The DBRs are dimensioned so to enhance the ATS response at $\lambda_C$, hence the TE-TE cavity is strongly over-coupled and no distinctive peak is visible in the measured spectra inside the TE bandgap region around $\lambda_{TE}$. On the other hand, a narrowband dip is apparent in the TE and TM outputs at $\lambda_{TM}$, revealing an interplay between the TE and TM modes at this working point, despite the offset from the polarization-converting resonance ($\lambda_C < \lambda_{TM}$). By tuning the length of the DBR, Fano-interference (FI) features in the spectrum can be measured in the polarization-converted output, as discussed in the next section.

## Photonic molecule: Fano interference regime

FI arises from the interference between a discrete state, with a fast phase variation, and a continuum state, with a slow phase variation. In our device architecture, as exemplified by *Figure 4*, the former corresponds to the TM-TM interaction and the latter to the TE-TM one in a DBR cavity operating at $\lambda_{TM}$. The relevant interactions are shown in *Fig. 4a*, where solid and dashed lines highlight phase matched and phase-mismatched processes, respectively.

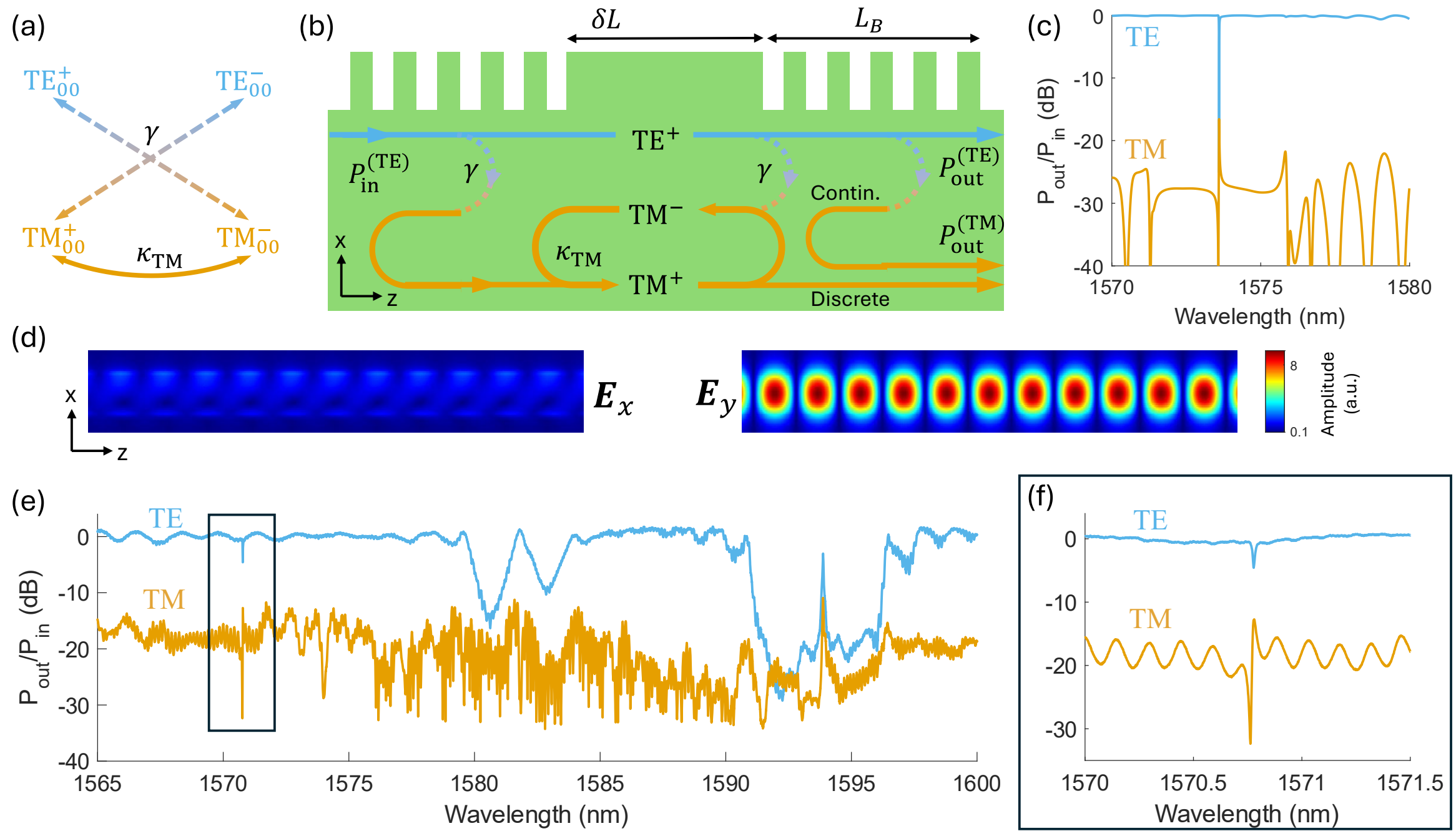

Figure 4. Polarization-hybrid photonic-molecule in the Fano resonance regime. (a) Modal coupling diagram at $\lambda_{TM}$. Solid arrows: phase-matched TM-TM coupling ($\kappa_{TM}$). Dashed arrows: phase-mismatched TE-TM coupling ($\gamma$). (b) Schematic of the relevant interactions a cavity and DBRs of lengths $\delta L$ and $L_B$, respectively, excited with a power $P_{in}^{(TE)}$ in the forward-propagating $TE_{00}$ mode. (c) TE (blue) and TM (orange) output powers, normalized to $P_{in}^{(TE)}$), obtained by CMT analysis for the DBR of Fig. 2 with $\delta L = \Lambda$, $L_B = 1000\,\Lambda$. (d) Propagation plots ($z$-$x$) of the TE and TM vectorial field components $E_x$ and $E_y$ at the center of a 100 µm-long cavity, at resonance. (e) Spectral distributions of the output TE (blue) and TM (orange) components measured for a cavity with $L_B = 1000\Lambda$ and $\delta L = \Lambda$. (f) Zoomed-in view of the TM-TM resonances around $\lambda_{TM}$.

With a $TE^+$ input wave, as shown in *Fig. 4b*, FI will be observed in the $TM^+$ device output, which originates from a $TE^+ \rightarrow TM^- \rightarrow TM^+$ energy transfer induced by the cascaded TE-TM and TM-TM interactions in the DBRs. When occurring in the first DBR, the above processes feed the cavity TM resonance (discrete state), while in the second DBR they directly convert the free-propagating $TE^+$ mode into an additional contribution (continuum) to the overall $TM^+$ device output. The interference of the two contributions is controlled by the grating length $L_B$, so that the transmitted spectra can be fine-tuned to exhibit a Fano-like response, as illustrated by the CMT simulations of *Fig. 4c*.

The device acts as a cavity (TM channel) side-coupled to a bus waveguide (TE channel) where the coupling is mediated by $\gamma$ and $\Delta\beta_C$. Therefore, the output spectra at the resonance wavelength $\lambda_{TM}$ exhibit a dip in the TE component and an asymmetric Fano line-shape in the TM component. A complementary situation occurs at the TE resonant wavelength $\lambda_{TE}$ and its detailed analysis is presented in *Supplementary Information Figure S7*. In this case, the TE transmission spectrum is similar to the one of a typical Bragg grating cavity,[35] with a narrowband resonance peak located at the center of the photonic bandgap, while the TM output exhibits a strongly asymmetric peak resulting from a dual pathway for TE-to-TM forward power flow.

Similarly to the ATS case of *Fig. 3c*, we performed also systematic simulations of integrated LNOI cavity devices for FI designs at $\lambda_{TE}$ and $\lambda_{TM}$, with independent numerical methods (Eigenmode Expansion solver, Ansys Lumerical) obtaining excellent agreement with the CMT results. The transmission spectra are compared in *Supplementary Information S4*. The computed intra-cavity electric field distributions for the two polarization components with a TE-excitation at $\lambda_{TM}$ ($\lambda_{TE}$) are shown in *Fig. 4d* (*Fig. S7d in Supplementary Information*). A strong standing-wave pattern in the TM component ($\boldsymbol{E_y}$) is apparent, as expected for operation at the resonance wavelength $\lambda_{TM}$, in *Fig. 4d*. At the same time the TE ($\boldsymbol{E_x}$) field distribution is essentially constant throughout the cavity. Only a very weak standing-wave pattern, resembling the $\boldsymbol{E_x}$-component of the resonating TM mode, can be discerned, consistently with the weak nature of the TE-TM grating coupling in this case.

The theoretical predictions were fully confirmed by the LNOI experiments as illustrated by *Fig. 4e*, showing the measured TE and TM output spectra from an integrated cavity with the same design as in *Fig. 3e*. To reveal the TM-TM resonance peak, the length of the DBRs on both sides of the cavity is reduced to $L_B$ = 1000$\Lambda$ = 435 µm. *Fig. 4f* shows a zoomed-in view of the measurements around $\lambda_{TM}$ = 1571 nm, for a comparison with *Fig. 4c*, making apparent the asymmetric line-shape in the TM output and the matching dip in TE transmission. Around $\lambda_{TE}$ = 1593 nm, we could also observe an asymmetric lineshape in the TM output, matching the theoretical analysis presented in *Supplementary Information S5.*

## Discussion

In conclusion, we have demonstrated integrated DBR polarization converters and Fabry–Pérot cavities operating in the telecom band and implemented on the LNOI photonic platform. The DBRs afford polarization conversion efficiencies of 86.5% within a compact footprint of 391.5 μm. The conversion mechanism relies primarily on strong optical confinement, without requiring higher-order mode engineering or anisotropic material responses,[25,29] thereby establishing a platform that is broadly compatible with a wide range of PIC material platforms. The performance can be further enhanced through straightforward geometric scaling, underscoring the flexibility and scalability of the approach.

One-dimensional cavities supporting hybrid TE-TM resonances revealed a rich landscape of interaction regimes engineerable by DBR design, which were experimentally explored considering the prototypical cases of strong coupling and Fano interference. The observation of ATS with a resonance separation of 120 pm and narrow linewidths of 16 pm confirmed operation in the strong coupling regime. Notably, although the cavity adopts a Fabry–Pérot geometry, the hybrid polarization states exhibit key characteristics of traveling-wave resonances (in contrast to Ref. [22,23]), reminiscent of clockwise and counterclockwise modal dynamics in ring resonators.[39,40] The observation in the cavity transmission spectra of strongly asymmetric Fano-like features, realized with alternative DBR designs, highlighted also the capability of the platform to control bound state in the continuum dynamics.

Overall, these results establish a versatile and highly engineerable platform that expands the accessible parameter space of integrated photonics, enabling advanced polarization control, modal hybridization, and tailored coupling and interference in one-dimensional structures. This provides a foundation for exploring fundamental phenomena such as quantum optical effects, coherent energy transfer, and parity–time symmetry in increasingly sophisticated designs and accessing the synthetic dimension provided by photon polarization in PIC devices. The intrinsic electro-optic and all-optical functionalities of the LNOI platform chosen for these first proof-of-principle device realizations, further enable dynamic control over interaction regimes. These capabilities open pathways to reconfigurable photonic devices, enhanced nonlinear interactions, and emerging topological and non-Hermitian platforms, with broad implications for optical signal processing, quantum information, and next-generation photonic technologies.

# Methods

## Device fabrication

We used 500 nm thick x-cut thin film lithium niobate on 2 μm $SiO_2$ wafer (NANOLN). The device designs were exposed via electron-beam lithography (Raith Voyager EBL System). The pattern is transferred into LNOI via argon-milling (AJA International, Inc. ATC Orion Series Ion Milling Systems) using the e-beam resist (ma-N 2405, micro resist technology GmbH) as mask. The redeposition from the etching is removed via base piranha solution ($NH_4OH$ : $H_2O_2$ = 3 : 1). Then, the LNOI chip is cladded with 10 μm thick photoresist (Epocore, micro resist technology GmbH) that is transparent in the NIR range. To make the waveguide facet for the edge coupling, the photoresist is exposed (MLA 150, Heidelberg Instruments) to open two trenches on top of the edge couplers. The facets are defined via another argon-milling session to etch through the 500 nm LNOI, using the resist as mask. Finally, the oxide layer is dry-etched (Oxford PlasmaPro 100 Cobra), before we cleave the silicon handle underneath. The resist is thick enough so that the rough surface on the top due to etching is well-separated from the waveguides.

## Optical measurements

Transmission spectra are recorded by sweeping the wavelength of a continuous-wave tunable laser (Santec TSL-570), while controlling and analyzing the device input and output polarization in free space via two linear polarizers, set to align with either the horizontal ($x$) or vertical ($y$) waveguide axes to control which polarization (TE or TM) to excite at the input and measure at the output. When rotating the input polarizer, a collimator and the polarization-maintaining (PM) fiber delivering the source signal are also rotated to always have the polarization of the source and the polarizer aligned. The light is coupled in and out of the chip via objectives (Mitutoyo MY50X-825). Finally, the transmitted light is measured by a power meter (Santec MPM-211).

# Data availability

The authors declare that relevant data supporting the findings of this study are available within the paper and its Supplementary Information. Additional data is available from the corresponding author upon reasonable request.

## Acknowledgements

The authors gratefully acknowledge support from the Knut and Alice Wallenberg Foundation through the Wallenberg Center for Quantum Technology (WACQT) and the Swedish Research Council (VR) (through grant No. 2016-06122) as well as scientific discussions and technical support from Prof. Vladislav Korenivski, Dr. Erik Holmgren and Dr. Adrian Iovan at the Albanova NanoLab MyFab facilities in Stockholm

## Author contributions

T. L. and K.G. conceived the idea. T.L. developed the theoretical analysis, fabricated the samples, and performed the experiments. Both authors analyzed the data, discussed the results and wrote the manuscript together.

## Competing interests

The authors declare no competing interests.

## Additional information

The online supplementary material contains information on: the models developed for the design and analysis of integrated DBR and cavity devices, including modelling of fabrication imperfections; numerical simulations performed independently by coupled mode theory and vectorial eigenmode expansion methods, and comparison of their results for the devices under test; further details on theory and experiments concerning the complementary Fano interference effect observed in the TE-TE bandgap for the device of Figure 4.